\documentclass{elsart}
\usepackage{natbib}
\usepackage{epsfig}
\usepackage{amssymb}

\graphicspath{{figsmall/}}

\begin{document}

\begin{frontmatter}
\title{ The WASA Detector Facility at CELSIUS}
\collab{CELSIUS/WASA Collaboration}
\runauthor{Chr.~Bargholtz et al.}
\author[Stock]{Chr.~Bargholtz},
\author[Tuebingen]{M.~Bashkanov},
\author[SINS]{M.~Ber{\l}owski},
\author[Novosibirsk]{A.~Bondar},
\author[Dubna]{D.~Bogoslawsky},
\author[Tuebingen]{W.~Brodowski},
\author[UCLA]{J.~Brudvik},
\author[UU]{H.~Cal\'en},
\author[UU]{F.~Capellaro},
\author[Novosibirsk]{A.~Chilingarov},
\author[Tuebingen]{H.~Clement},
\author[Arizona]{J.~Comfort},
\author[Hamburg]{L.~Demir\"ors},
\author[TSL]{C.~Ekstr\"om},
\author[UU]{K.~Fransson},
\author[TSL]{C.-J.~Frid\'en},
\author[Stock]{L.~Ger\'en},
\author[MEPI]{M.~Gornov},
\author[MEPI]{V.~Grebenev},
\author[Hamburg]{J.~Greiff},
\author[MEPI]{Y.~Gurov},
\author[UU]{L.~Gustafsson},
\author[UU]{B.~H\"oistad},
\author[Dubna]{G.~Ivanov},
\author[UU]{M.~Jacewicz},
\author[Dubna]{E.~Jiganov},
\author[UU]{A.~Johansson},
\author[UU]{T.~Johansson},
\author[UU]{S.~Keleta},
\author[Tuebingen]{O.~Khakimova},
\author[Muenst]{A.~Khoukaz},
\author[Juelich]{K.~Kilian},
\author[Tsukuba]{N.~Kimura},
\author[UU]{I.~Koch},
\author[Novosibirsk]{G.~Kolachev},
\author[Tuebingen]{F.~Kren},
\author[UU]{S.~Kullander},
\author[UU]{A.~Kup\'s\'c},
\author[Novosibirsk]{A.~Kuzmin},
\author[Dubna]{A.~Kuznetsov},
\author[Stock]{K.~Lindberg},
\author[UU]{P.~Marciniewski},
\author[Tuebingen]{R.~Meier},
\author[Tuebingen]{O.~Messner},
\author[Dubna]{B.~Morosov},
\author[SINS]{A.~Nawrot},
\author[UCLA]{B.M.K.~Nefkens},
\author[TSL]{G.~Norman},
\author[Juelich]{W.~Oelert},
\author[Hamburg]{C.~Pauly},
\author[UU]{H.~Pettersson},
\author[Novosibirsk]{A.~Pivovarov},
\author[Tuebingen]{J.~P\"atzold},
\author[Dubna]{Y.~Petukhov},
\author[Dubna]{A.~Povtorejko},
\author[TSL]{D.~Reistad},
\author[UU]{R.J.M.Y.~Ruber},
\author[Dubna]{S.~Sandukovsky},
\author[UU]{K.~Sch\"onning},
\author[Hamburg]{W.~Scobel},
\author[Juelich]{T.~Sefzick},
\author[MEPI]{R.~Shafigullin},
\author[Novosibirsk]{B.~Shwartz},
\author[Novosibirsk]{V.~Sidorov},
\author[Tuebingen]{T.~Skorodko},
\author[ITEP]{V.~Sopov},
\author[UCLA]{A.~Starostin},
\author[SINS]{J.~Stepaniak},
\author[Novosibirsk]{A.~Sukhanov},
\author[Dubna]{A.~Sukhanov},
\author[ITEP]{V.~Tchernyshev},
\author[Stock]{P.-E.~Tegn\'er},
\author[UU]{P.~Th\"orngren Engblom},
\author[Dubna]{V.~Tikhomirov},
\author[Osaka]{H.~Toki},
\author[WU]{A.~Turowiecki},
\author[Tuebingen]{G.J.~Wagner},
\author[UU]{U.~Wiedner},
\author[WU]{Z.~Wilhelmi},
\author[Juelich]{M.~Wolke},
\author[Tsukuba]{A.~Yamamoto},
\author[Tsukuba]{H.~Yamaoka},
\author[SINS]{J.~Zabierowski},
\author[Stock]{I.~Zartova},
\author[UU]{J.~Z{\l}oma\'nczuk}
\address[Stock]{Stockholm University, Stockholm, Sweden}
\address[Tuebingen]{Physikalisches Institut der Universit\"at T\"ubingen,
T\"ubingen, Germany}
\address[SINS]{Soltan Institute of Nuclear Studies, Warsaw and Lodz, Poland}
\address[Novosibirsk]{Budker Institute of Nuclear Physics, Novosibirsk, Russia}
\address[Dubna]{Joint Institute for Nuclear Research, Dubna, Russia}
\address[UCLA]{University of California Los Angeles, Los Angeles, 
California, USA}
\address[UU]{Uppsala University, Uppsala, Sweden}
\address[Arizona]{Arizona State University, Tempe, USA}
\address[Hamburg]{Institut f\"ur Experimentalphysik der Universit\"at Hamburg,
Hamburg, Germany}
\address[TSL]{The Svedberg Laboratory, Uppsala, Sweden}
\address[MEPI]{Moscow Engineering Physics Institute, Moscow, Russia}
\address[Muenst]{ M\"unster University, M\"unster, Germany}
\address[Juelich]{Institut f\"ur Kernphysik, Forschungszentrum J\"ulich,
J\"ulich, Germany}
\address[Tsukuba]{High Energy Accelerator Research Organization,
Tsukuba, Japan}
\address[ITEP]{Institute of Theoretical and Experimental Physics,
Moscow, Russia}
\address[Osaka]{Research Centre for Nuclear Physics, Osaka, Japan}
\address[WU]{Institute of Experimental Physics of Warsaw University, Warsaw, Poland}

\begin{abstract}

The  WASA 4$\pi$  multidetector system,  aimed at  investigating light
meson production in light ion  collisions and $\eta$ meson rare decays
at  the CELSIUS  storage ring  in  Uppsala is  presented.  A  detailed
description of the design,  together with the anticipated and achieved
performance parameters are given.

\end{abstract}

\begin{keyword}
 \PACS 29.40.-n \sep  29.40.Mc \sep  29.40.Cs \sep 13.20.-v \sep 14.40.Aq
\end{keyword}

\end{frontmatter}

\section{Introduction}

The  4$\pi$  detector  facility  WASA\footnote{CELSIUS/WASA  Homepage,
http://www.tsl.uu.se/wasa} was designed  for studies of production and
decays of light mesons in an internal-target experiment at the CELSIUS
accelerator and cooler  storage ring ~\cite{Holm:1986jq}.  The highest
beam-proton  kinetic  energy reachable  at  CELSIUS  was  1.5 GeV  and
protons of  energies up to  500~MeV could be electron  cooled.  Single
and  multipion   production  in  proton-proton,   proton-deuteron  and
deuteron-deuteron interactions were studied  over a wide range of beam
energies.  The available beam  energies also allowed production of eta
mesons in  proton-proton and proton-deuteron  reactions and production
of omega mesons in proton-deuteron reactions.

The WASA  project required some special technical  developments e.g. a
very  thin-walled  superconducting  solenoid  and  a  hydrogen  pellet
target.  The  pellet concept is crucial  to achieve a  close to 4$\pi$
detection acceptance in this internal-target storage ring experiment.

The WASA  detector was installed at  CELSIUS in 1999 and  from 2002 it
was  fully equipped  with read-out  electronics. The  WASA experiments
were assigned in  total about 6500 hours of beam  time until June 2005
when  CELSIUS  was  closed down.   WASA  was  then  moved to  COSY  at
Forschungzentrum  J\"ulich,  Germany,  where it  successfully  started
operation again in the autumn of 2006.

In  the  CELSIUS/WASA  experiment  several production  reactions  were
studied  e.g.:  $pd\to^{3}$He$\pi\pi$ \cite{Bashkanov:2005fh},  $pp\to
pp3\pi$                  ~\cite{Pauly:2006pm},                  $pp\to
pp\pi^0$~\cite{ThorngrenEngblom:2006dx},        $pd\to^{3}$He~$\omega$,
$dd\to^{4}$He~$\pi\pi$.     Some   $\eta$   decays    e.g.    $\eta\to
3\pi^0$~\cite{Bashkanov:2007iy},
$\eta\to\pi^+\pi^-\e^+\e^-$~\cite{Bargholtz:2006gz}                 and
$\eta\to\e^+\e^-\gamma$~\cite{Berlowski:2007zx}    were   studied   in
$pd\to^{3}$He$\eta$  and $pp\to pp\eta$  reactions.  New  upper limits
for  the  $\eta\to\e^+\e^-\e^+\e^-$ and  $\eta\to\mu^+\mu^-\mu^+\mu^-$
decay branching ratios~\cite{Berlowski:2007zx} have been obtained.

\section{The WASA detector}
\label{sec:setup}
\label{subwasa}

\begin{figure}[ht]
\includegraphics[width=\textwidth,height=0.4\textheight,clip]{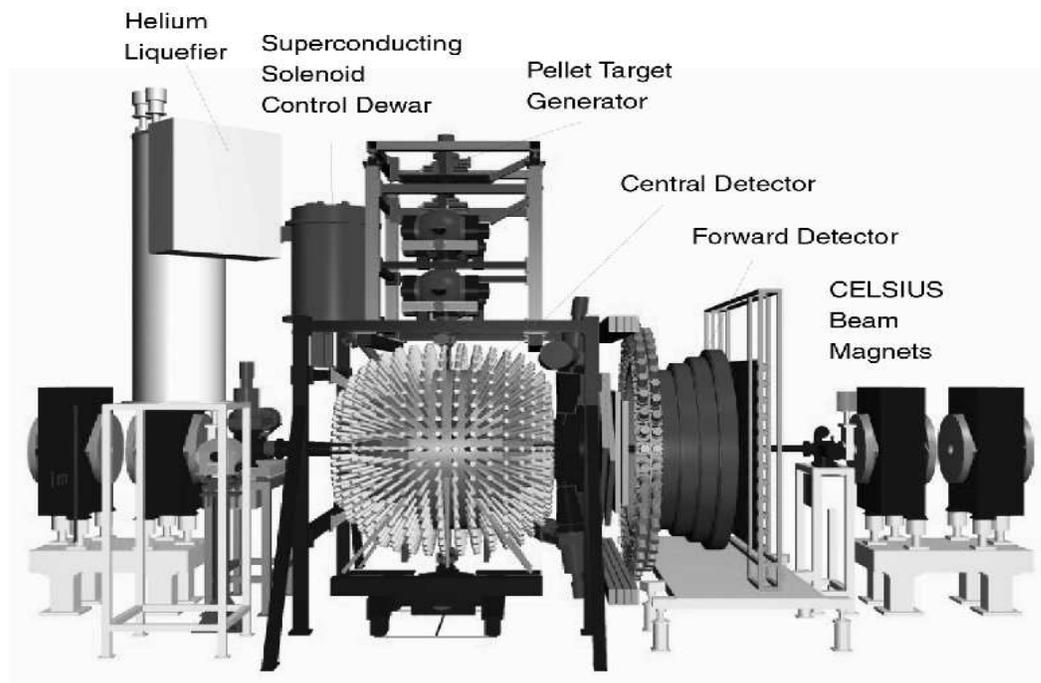}
\caption{  \label{CAD} CAD  view of  the WASA  Detector  Facility. The
zero-degree spectrometer is located further downstream to the right.}
\end{figure}

The  WASA setup at  CELSIUS, shown  in the  form of  a CAD  drawing in
Fig.~\ref{CAD},   can  be   divided   into  four   major  parts:   the
pellet-target, the forward detector (FD), the zero-degree spectrometer
(ZD) and the central detector (CD).

The pellet-target  system is integrated  in the setup and  it provides
small   spheres  of   frozen   hydrogen  or   deuterium  as   internal
targets. This  allows high luminosity and high  detection coverage for
meson decay products like photons, electrons and charged pions.

\begin{figure}[!htb]
\includegraphics[width=\textwidth,clip]
{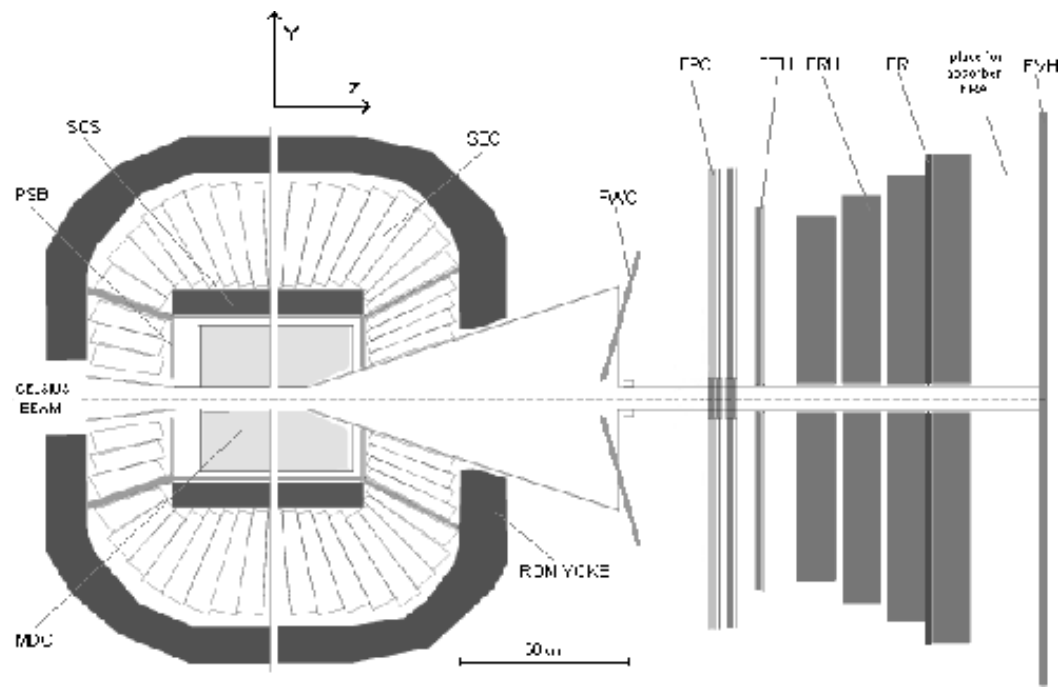}
\caption{\label{fig:wasa}  Cross  section of  the  WASA detector.  The
central detector built  around the interaction point (at  the left) is
surrounded by  an iron  yoke. The layers  of the forward  detector are
visible  on  the  right-hand   side.  The  individual  components  are
described in the text.}
\end{figure}

The FD  and ZD measure  charged target-recoil particles  and scattered
projectiles.    The  FD   consists   of  eleven   planes  of   plastic
scintillators  and  of  proportional  counter  drift  tubes.   The  ZD
contains  detectors placed in  the CELSIUS  beam magnets  and provides
information on  forward going particles, that would  otherwise be lost
in the  beam pipe. This was  exploited e.g. for  the identification of
pd$\to^{3}$He$\eta$ events measured  close to the kinematic threshold.
The CD was  designed for measurements of the  meson decay products and
consists  of  an   electromagnetic  calorimeter  of  CsI(Na)  crystals
surrounding  a superconducting  solenoid.   Inside of  the solenoid  a
cylindrical  chamber   of  drift  tubes   and  a  barrel   of  plastic
scintillators  are  placed.  A  vertical  cross  section  of the  WASA
detector is shown in Fig.~\ref{fig:wasa}.

\subsection{Pellet target}
\label{sec:pellet}

\begin{figure}[!htbp]
\centering
\includegraphics[width=\textwidth]{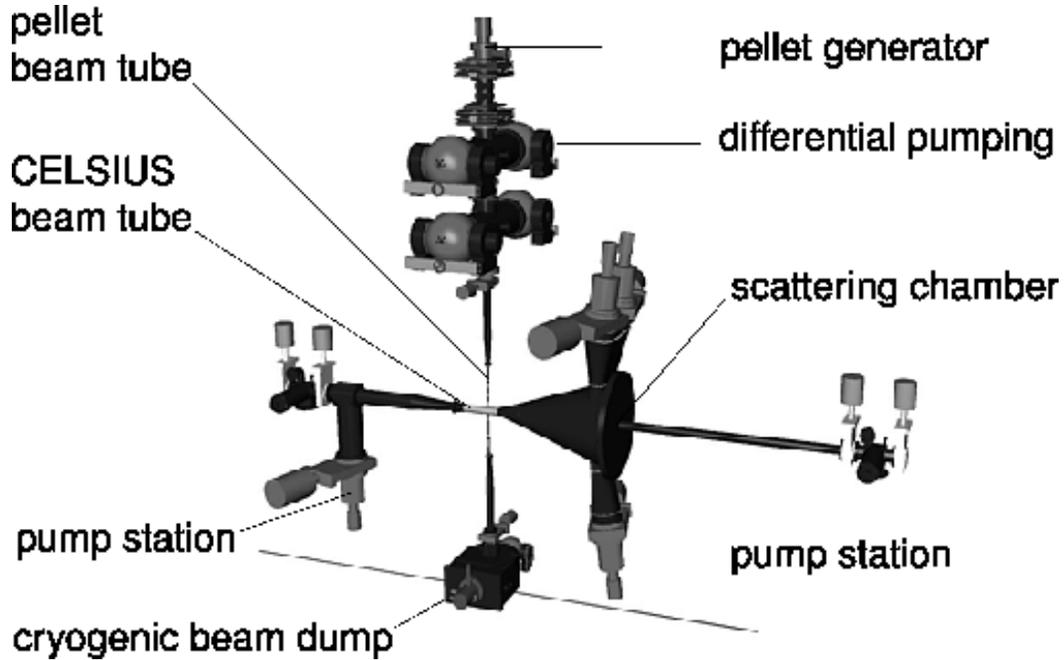}
\caption{\label{fig:pellet1}Layout of the pellet target system.}
\end{figure}

The pellet target system was a unique development for the CELSIUS/WASA
experiments     \cite{Trostell:1995in,Ekstrom:1996jt}.     The    main
components  of the  system are  shown in  Fig.~\ref{fig:pellet1}.  The
heart  of the  setup is  the pellet  generator where  a jet  of liquid
hydrogen is broken up into droplets with a diameter of about 35~$\mu$m
by a vibrating nozzle. The droplets freeze by evaporation in a droplet
chamber  and   form  a  beam  of   pellets  that  pass   a  7~cm  long
vacuum-injection  capillary.   After   collimation,  the  pellets  are
directed through a thin 2~m  long pipe into the scattering chamber and
further down to a pellet beam dump.  The inner diameter of the pipe is
5~mm  at the  entrance to  the scattering  chamber.   This arrangement
provides the necessary space to put the 4$\pi$ detection system around
the interaction  region.  Pellet target  thicknesses of up  to $3\cdot
10^{15}$~atoms/cm$^2$  gave acceptable  half-lives of  the circulating
ion  beam  as  well  as  acceptable vacuum  conditions.  Some  of  the
parameters of the pellet  target are listed in table \ref{tab:pellet}.
The pellet target system was regularly operated with pellets of normal
hydrogen or deuterium.

\begin{table}[!htbp]
\centering
\small
\begin{tabular}{lr}
\hline
Pellet diameter [$\mu$m]       &  25 - 35     \\
Pellet frequency [kHz]        &  5 - 12  \\
Pellet - pellet distance [mm]     &  9 - 20   \\
Effective target thickness [at.s/cm$^2$] &  $> 10^{15}$ \\
Beam diameter [mm]             & $2 - 4$   \\
 \hline
\end{tabular}
\normalsize
\caption{\label{tab:pellet} Performance of the pellet target system.}
\end{table}

\subsection{Forward Detector}

The  FD  was  designed  mainly  for detection  and  identification  of
scattered  projectiles  and  charged  recoil particles  like  protons,
deuterons  and He  nuclei in  $\pi$ and  $\eta$  production reactions.
Also  neutrons and  charged pions  can  be measured.   All FD  plastic
scintillators  may  supply information  for  the  first level  trigger
logic.  Most of the components of  this part of the setup were used in
a previous experiment at  CELSIUS, PROMICE-WASA, which is described in
more  detail  in  ref.~\cite{Calen:1996ft}.   A summary  of  the  most
important  features  of  the   forward  detector  is  given  in  table
\ref{tab:FD}.  In this section the individual components are described
in some  detail with  emphasis on the  extensions with respect  to the
PROMICE-WASA setup .

\begin{table}[!h]
\centering
\begin{tabular}{lr}
\hline
Total number of scintillator elements  & 280\\
Scattering angle coverage [degrees] & 3 - 17 \\
Scattering angle resolution [degrees] & 0.2\\
Amount of sensitive material [g/cm$^{2}$] & 50\\ 
~~~~[radiation lengths] & $\approx$ 1 \\ 
~~~~[nuclear interaction lengths] & $\approx$ 0.6 \\
Thickness of vacuum window (st. steel) [mm] & $\approx$ 0.4\\ 
Maximum kinetic energy ($T_{stop}$) for stopping: & \\
~~~~$\pi^{\pm}$/proton/deuteron/alpha\qquad [MeV] & 170/300/400/900\\
Time resolution [ns]                    & $<$ 3\\
Energy resolution for: & \\
~~~~ stopped particles & $\approx$ $3\%$\\
~~~~ particles with $T_{stop}$ $<T<$ 2$T_{stop}$ & 4 - $8\%$\\ 
Particle identification & $\Delta E-E$\\ 
\hline
\end{tabular}
\normalsize
\caption{\label{tab:FD} Some features of the Forward Detector.}

\end{table}

\begin{figure}[ht]
\hspace{3cm}\includegraphics[width=0.5\textwidth,clip]{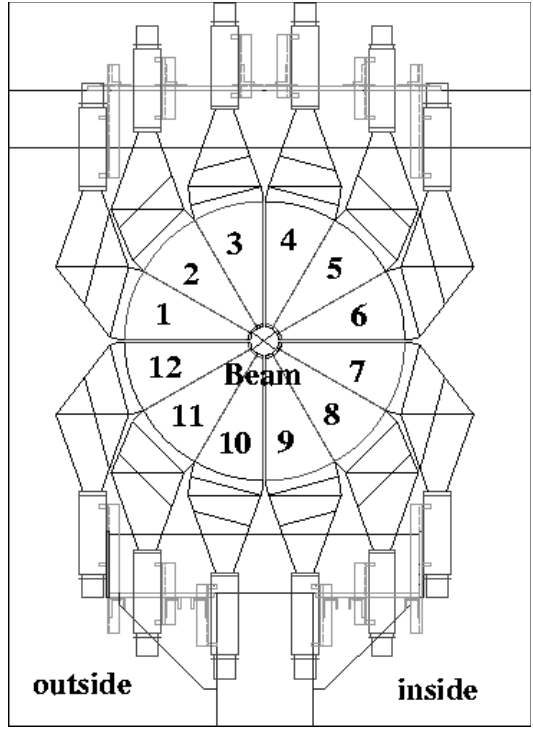}
\caption{\label{fig:fwc} Schematic view of the FWC.}
\end{figure}

The Forward Window Counter (FWC) is the first detector layer in the FD
(along the beam direction) and consists of 12 plastic scintillators of
5~mm  thickness (Fig.~\ref{fig:fwc}).   It is  mounted tightly  on the
paraboloidal stainless  steel vacuum window.   Therefore, the elements
are inclined with approximately 19$^{\circ}$ with respect to the plane
perpendicular to the beam direction.   The FWC signals are used in the
first level trigger logic to reduce the background caused by particles
scattered  in  the downstream  beam  pipe and  in  the  flange at  the
entrance to the  FD. The signals are also used  to select He ejectiles
at the trigger level.

Downstream  the   FWC,  there  is  a  tracking   device:  the  Forward
Proportional  Chambers (FPC).  It  is composed  of four  modules, each
with four staggered layers of  122 proportional drift tubes (so called
straws)  of  8~mm diameter.   The  straws,  made  of 26  $\mu$m  thick
aluminized Mylar and with a central sense wire of stainless steel with
35~$\mu$m diameter,  are operated with a 50/50  $Ar/CO_2$ gas mixture.
The modules are  rotated by 45$^\circ$ with respect  to each other (in
the plane perpendicular to the beam axis).  They are used for accurate
reconstruction  of  track  coordinates  and  provide  precise  angular
information of  the particles originating from the  target region. The
FPC  is described  in detail  in ref.~\cite{Calen:1996ft}  and  in the
Ph.D. thesis of J.~Dyring~\cite{Dyring:1997aa}.

The  Forward Trigger Hodoscope  (FTH), consisting  of three  layers of
5~mm thick  plastic scintillators, is  placed next to the  FPC.  There
are 24 Archimedean spiral-shaped elements  in the first two planes and
48 radial elements in the  third.  The special geometry, combining all
three  layers, results  in  a  pixel structure,  which  is useful  for
resolving  multi-hit  ambiguities~\cite{Dahmen:1994bd}.   The FTH  was
designed to  provide scattering angle  information in the  first level
trigger logic.

\begin{figure}[!hbt]
\centering
\includegraphics[width=0.7\textwidth,clip]{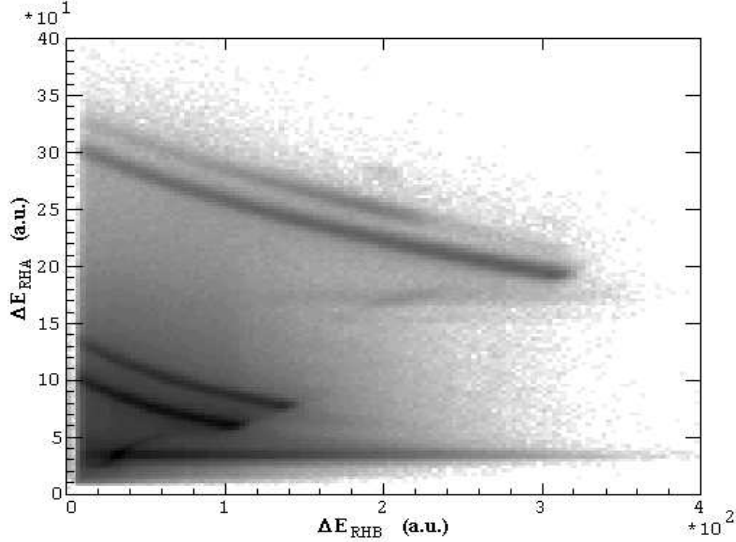}
\caption{\label{fig:FDdep}   An  example   of   $\Delta  E-\Delta   E$
information from a deuteron-deuteron run at 1012~MeV used for particle
identification in the forward detector.  Energy deposited in the first
layer  of FRH is  plotted versus  the energy  deposited in  the second
layer  of FRH.  The  upper two  bands are  from $^{4}$He  and $^{3}$He
nuclei and  in the  lower part of  the plot  there are two  bands from
deuterons and protons. }
\end{figure}

Behind the FTH,  the four layers of the  Forward Range Hodoscope (FRH)
are positioned.  Each plane is made of 24 plastic scintillator modules
each  of 11~cm  thickness. The  FRH, together  with FTH,  is  used for
energy   determination   of  charged   particles   and  for   particle
identification by $\Delta  E-E$ technique.  Fig.~\ref{fig:FDdep} shows
how protons, deuterons and He nuclei can be identified.

The  identity and  initial kinetic  energy  of a  charged particle  is
reconstructed from  the pattern of  deposited energy in  the different
detector planes.  Even  for particles that are stopped  in a detector,
the  total deposited  energy  is different  from  the initial  kinetic
energy.   This is  because  some of  the  energy is  lost in  inactive
material  between the  detectors elements,  in the  scattering chamber
window, etc.  The  variation of the deposited energy  is strong enough
to be useful for energy  reconstruction also for high energy particles
not stopped in  the detector material.  For protons,  this can be used
in the kinetic energy range 300~MeV  to 800~MeV and for deuterons in a
similar  energy  range   starting  from  400~MeV.   Identification  of
punch-through  particles can  be done  using either  the  forward veto
hodoscope  or the  $\Delta E$  information from  the last  FRH planes.
More  details  on  FTH,  FRH   and  the  method  for  particle  energy
reconstruction are given in ref.~\cite{Calen:1996ft}.

\begin{figure}[ht]
\hspace{3cm}\includegraphics[width=0.5\textwidth,clip]{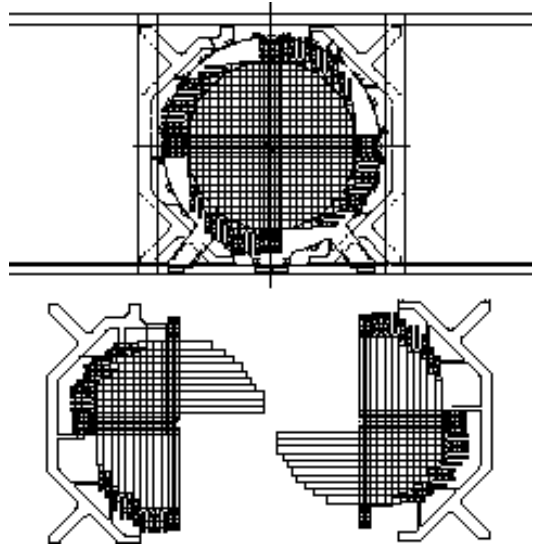}
\caption{\label{fig:fri}  Schematic view  of the  FRI  detector (upper
picture) and  its two  planes with orthogonally  oriented scintillator
bars (lower picture).}
\end{figure}

Between  the  third  and  fourth  layers  of the  FRH  there  are  two
interleaving  layers  of 5~mm  thick  plastic  scintillator bars  with
one-sided readout (Fig.~\ref{fig:fri}), the Forward Range Interleaving
Hodoscope (FRI).  Each layer has 32 bars, oriented horizontally in one
and vertically in the other.  The main purpose of this addition to the
FRH is  to provide a  two-dimensional position sensitivity  inside the
FRH necessary for measurement  of scattering angles for neutrons.  The
probability that  a neutron of a  few hundred MeV  kinetic energy will
interact in the FRH is around  35~\%.  The FRI can help to reconstruct
outgoing  protons from  the  nuclear interaction  and to  discriminate
against background  tracks due to  secondary interactions in  the beam
pipe and other structural material.  More information about the design
and     performance    of     the    FRI     can    be     found    in
references~\cite{Pauly:2006zz, Pauly:2005vn}.

\begin{figure}[ht]
\hspace{3cm}\includegraphics[width=0.5\textwidth,clip]
{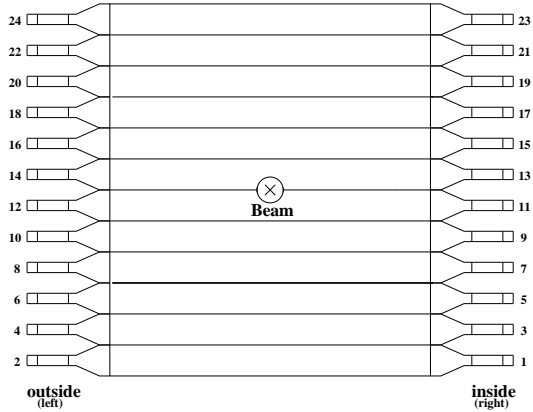}
\caption{\label{fig:fvh} Schematic view of the Forward Veto Hodoscope.}
\end{figure}

The last detector  layer of the FD is a  wall of plastic scintillators
(Fig.~\ref{fig:fvh}), the  Forward Veto Hodoscope  (FVH).  It consists
of   12   horizontal   plastic   scintillator  bars,   equipped   with
photomultipliers on  both ends.  The hit  position along a  bar may be
reconstructed from signal time  information with a resolution of about
5~cm ($\sigma$).  In the first  level trigger the signals are used for
rejection (or selection) of particles punching through the FRH.

Optionally,  a  passive absorber  layer  (FRA)  made  of iron  can  be
introduced  between  the last  layer  of the  FRH  and  the FVH.   The
thickness   of  the   absorber  can   be  chosen   from  5~mm   up  to
100~mm. Immediately in  front of and behind the  iron are placed 10~mm
thick plexi-glass sheets.  The absorber has been used for stopping the
protons  from the  $pp\to pp\eta$  reaction  at a  beam proton  energy
around  1360~MeV.   In  this  case  the faster  protons  from  elastic
scattering  and from  pion  production penetrate  the  FRA and  induce
signals in  the FVH which can be  used for veto purposes  in the first
level trigger.

\subsection{Zero-Degree spectrometer}

\begin{figure}[!hbt]
\centering
\centerline{\includegraphics[width=0.9\textwidth,clip]{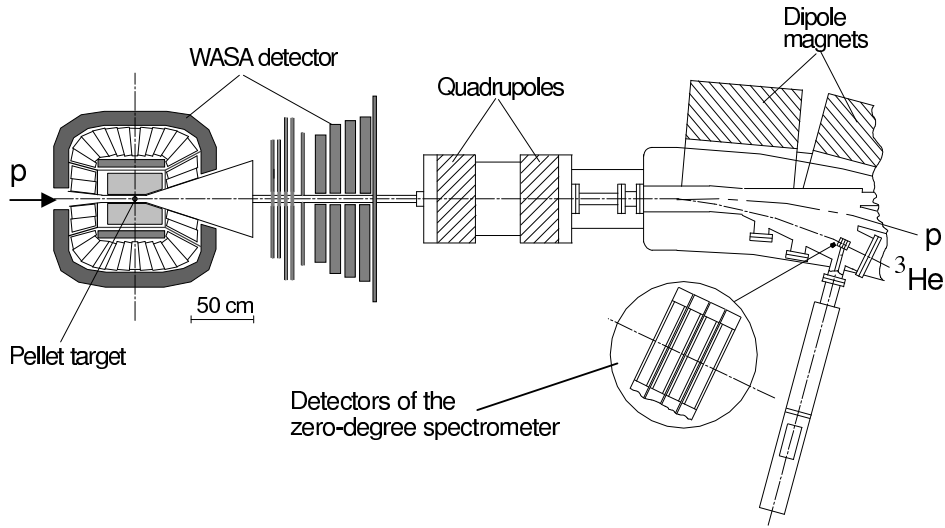}}
\caption{\label{fig:deeTSL} Drawing of  CELSIUS dipoles downstream the
WASA detector with the zero-degree detector installation. }
\end{figure}

The ZD uses the CELSIUS  dipole magnets for filtering out low momentum
reaction       products       escaping       into       the       beam
pipe~(figure~\ref{fig:deeTSL}).    The  reaction  products   could  be
detected  at different  port positions  of the  vacuum-chamber  in the
dipole  magnets  (about 6.6~m  away  from  the  target).  A  range  of
magnetic  rigidities   relative  to  that  of   the  circulating  beam
$(B\rho)_{rel}$  of  0.3 to  0.8  could  be  covered.  The  acceptance
depends critically on the size of the detectors and on the settings of
the quadrupole magnets.  The maximum geometric acceptance, 0.7~msr, is
reached at a  relative rigidity of 0.65.  A  stainless steel window of
100~$\mu$m thickness at the port  position allowed the use of external
setups of detectors.  Setups of  6 layers of 12 CsI(Tl) crystals, some
front Si-strip  detectors and plastic scintillators  were prepared for
this purpose.

\begin{figure}[!hbt]
\centering
\centerline{\includegraphics[width=0.9\textwidth,clip]{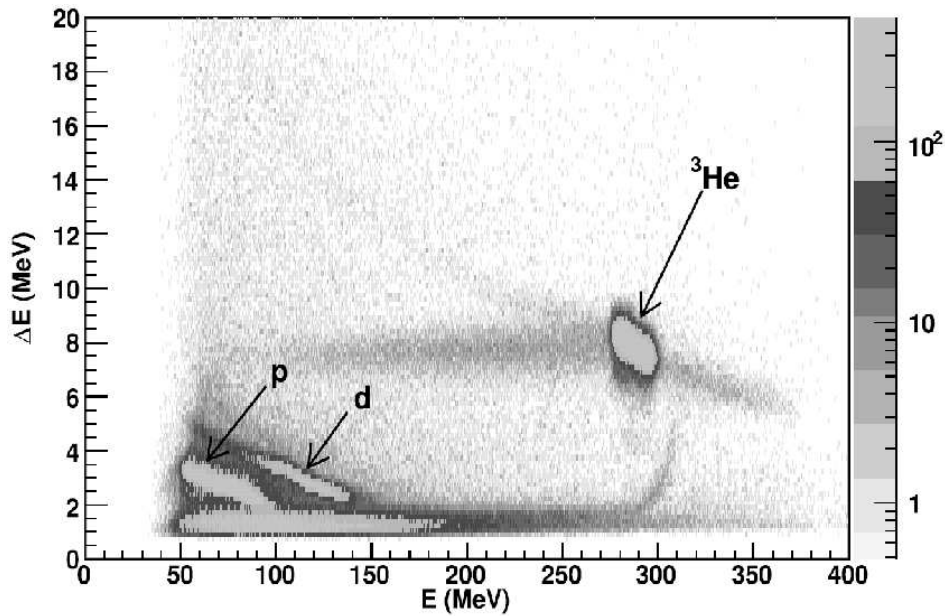}}
\caption{\label{fig:deeTSR}  An  example  of  particle  identification
using $\Delta E-E$ information from the zero-degree spectrometer for a
raw  data  sample of  $pd$  interactions  close to  $pd\to^3$He$\eta$
threshold. }
\end{figure}

The  main detector  setup was  a  telescope of  silicon and  germanium
detectors  cooled to liquid-nitrogen  temperature installed  inside of
the vacuum-chamber.   This was used  for very clean tagging  of $\eta$
production in  $pd\to^{3}$He$\eta$ reactions near  the threshold.  The
telescope  comprised  two  thin  silicon  and  two  thicker  germanium
detectors.  The silicon detectors  installed in front of the germanium
ones  are used  as  transmission ($\Delta  E$)  elements for  particle
identification.   The  $^{3}$He   ions  from  the  $pd\to^{3}$He$\eta$
reaction are stopped in the thicker germanium detectors. $^{3}$He ions
with an  energy of up to  approximately 400~MeV were  fully stopped in
the detector telescope (Fig.~\ref{fig:deeTSR}).  The energy resolution
(FWHM) of  the telescope under normal CELSIUS  conditions for $^{3}$He
ions with an energy of  approximately 300~MeV is estimated to be about
1.5  ~MeV. A  detailed description  of  this detector  system and  its
performance can be found in ref.~\cite{Bargholtz:2006rv}.

\subsection{Central Detector}

The CD is  built around the interaction point  and was designed mainly
for detection and identification of  the decay products of $\pi^0$ and
$\eta$ mesons: photons, electrons and charged pions. It consists of an
inner drift  chamber (MDC), thin  plastic scintillators in  a cylinder
geometry  (PSB),  a solenoid  (SCS)  providing  a  magnetic field  for
momentum  measurements and  a CsI  calorimeter (SEC).   The  amount of
structural material is kept minimal  to reduce the disturbances on the
particles.  The  beam pipe is made  of 1.2~mm thick  beryllium and the
total thickness of the  solenoid corresponds to 0.18 radiation lengths
only. The placement  of the solenoid inside of  the calorimeter allows
the use of photomultipliers for readout.

The main requirements for the design were the following:
\begin{itemize}
\item to  handle high particle  fluxes at  luminosities around
${\rm 10^{32}\,cm^{-2}s^{-1}}$.
\item to  measure photons with energies  from a few
MeV up to 800~MeV.
\item to  measure, in a  magnetic field of  about 1 T, the  momenta of
electrons and  positrons in  the range 20~MeV/c  to 600~MeV/c  with an
accuracy ${\sigma(p)/p \approx 2\%}$.
\end{itemize}

The momenta  of heavier  charged particles can  also be measured  in a
similar momentum range but with lower accuracy. For pions and muons an
accuracy ${\sigma(p)/p \approx 4\%}$  (for $p$ 100~MeV/c to 600~MeV/c)
can be obtained and for  protons a ${\sigma(p)/p\approx 6\%}$ (for $p$
200~MeV/c to 800~MeV/c).

The   main   components   of    the   Central   Detector,   shown   in
Fig.~\ref{fig:wasa}, are presented below in some detail.

\subsubsection{The Superconducting Solenoid - (SCS)}
The SCS provides an axial magnetic field for the momentum measurements
of  the tracks  measured by  the MDC.  It also  protects the  CD from
low-energy  delta  electrons  produced  in the  interactions  of  beam
particles  with  the  pellets.   The  wall thickness  of  the  SCS  is
minimized in order  to allow high accuracy of  the energy measurements
in the calorimeter.  The return path for the magnetic flux is provided
by a yoke made out of 5 tons of soft iron with low carbon content. The
yoke  shields the  readout  electronics from  the  magnetic field  and
also serves  as  support  for  the  calorimeter  crystals.  The  main
parameters of the SCS are given in table \ref{tab:scs}.

\begin{table}[!h]
\centering
\begin{tabular}{lr}
\hline
Superconducting coil&\\
~~~~ Inner/outer radius [mm] & 267.8 / 288.8\\ 
~~~~ Superconductor (stabilizer) & NbTi/Cu (pure Al)\\ 
~~~~ Total winding length [mm] & 465\\
~~~~ Maximum central magnetic flux density, {B$_{c}$} [T] & 1.3\\
~~~~ Field uniformity in the MDC [T] & 1.22 $\pm$$0.25$\\
~~~~ Cooling (Liquid He) [K] & 4.5 \\ 
Cryostat&\\
~~~~ Material & Aluminium\\
~~~~ Inner / outer radius [mm] & 245 / 325\\ 
~~~~ Overall length [mm] & 555\\ 
 SCS wall thickness (coil+cryostat) [radl] & {\bf 0.18}\\ 
\hline
\end{tabular}
\caption{\label{tab:scs} Main parameters of the superconducting
coil and its cryostat.}
\end{table}

In order to  map the magnetic field inside the  volume enclosed by the
SCS, the magnetic field strength inside the MDC was measured with Hall
probes and,  in addition, the  field distribution was  calculated with
simulation programs. The calculated values were fitted to the measured
ones with an  error of $\pm$1\% of B$_{total}$.   The SCS is described
in  detail in  ref.~\cite{Ruber:2003aa}  and in  the  Ph.D. thesis  of
R.~Ruber \cite{Ruber:1999zz}.

\subsubsection{The Mini Drift Chamber - (MDC)}
The MDC is  placed around the beam pipe and  is used for determination
of particle momenta  and reaction vertex. It is  a cylindrical chamber
covering scattering  angles from 24$^\circ$ to  159$^\circ$. For large
angle  scattered  protons  from  elastic proton-proton  scattering,  a
vertex resolution  ($\sigma$) of  0.2~mm perpendicular and  3~mm along
the beam axis  can be reached.  A detailed description  of the MDC can
be found in the Ph.D. thesis of M.~Jacewicz~\cite{Jacewicz:2004zz}.

\begin{figure}[!h]
\includegraphics[width=0.425\textwidth,]{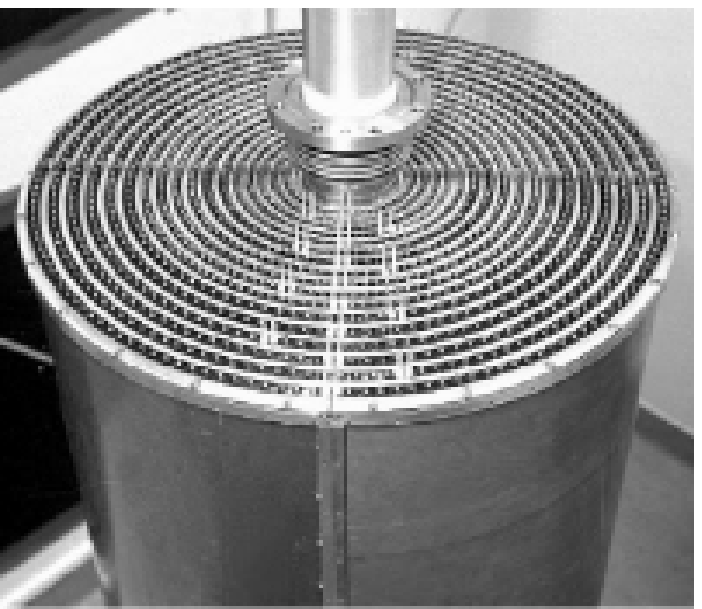}%
\includegraphics[width=0.575\textwidth,]{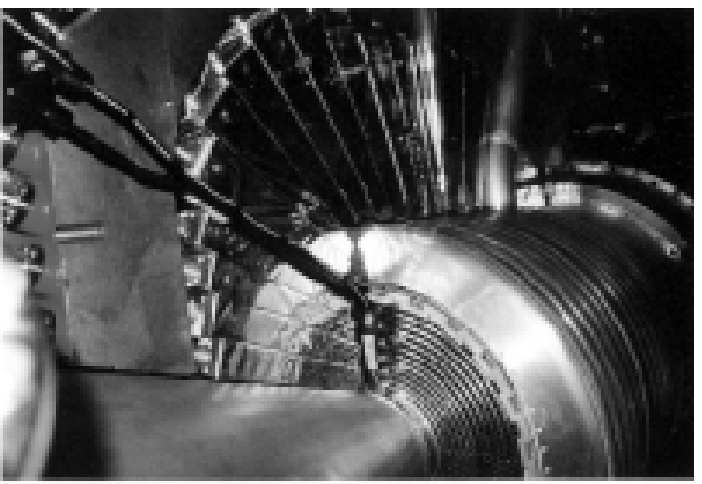}
\caption{\label{fig:mdccyl} (Left) The fully assembled MDC inside 
the Al-Be cylinder. (Right) The MDC surrounded by PSB elements 
and the SCS cryostat.}
\end{figure}

\begin{figure}[!hbt]
\centering
\includegraphics[width=0.6\textwidth]{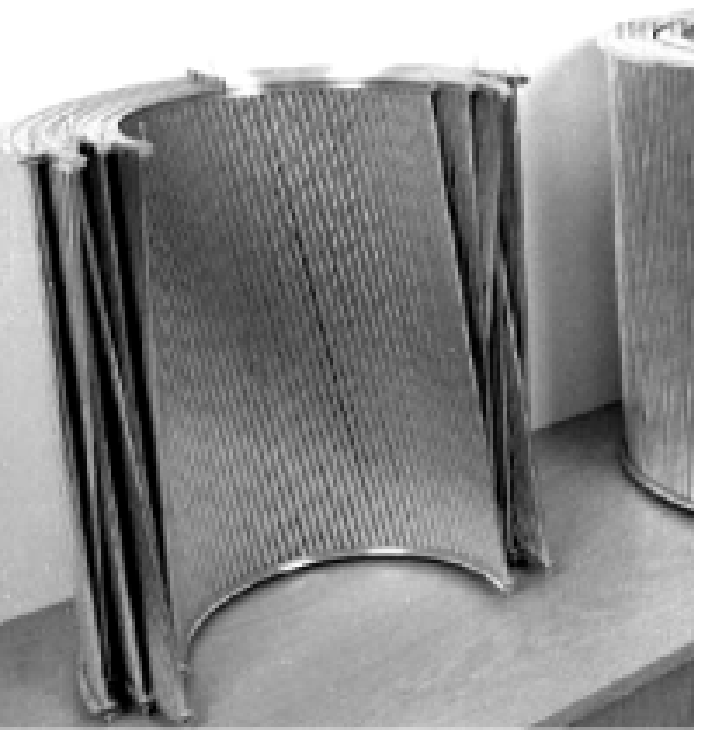}%
\includegraphics[width=0.4\textwidth]{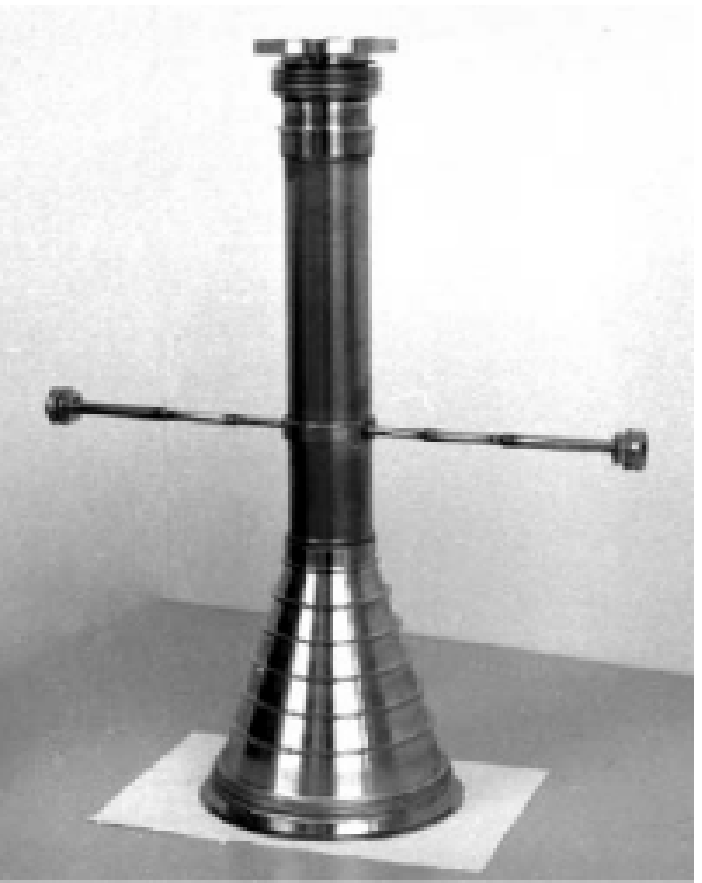}
\caption{\label{fig:mdcpipe}  (Left)   Drift  tubes  secured   in  the
end-plates.  Note the  {\em stereo}  layers interleaved  with parallel
layers. (Right)  Be beam  pipe with pellet  pipe crossing  and forward
cone.}
\end{figure}

The MDC consists of 1738 drift tubes, so called straws, arranged in 17
cylindrical layers.   The diameter  of the straws  in the  5 innermost
layers is  4~mm, 6~mm in  the next  6 layers and  8~mm in the  outer 6
layers.  The straws  are made of a thin  (25~$\mu$m) mylar foil coated
with 0.1~$\mu$m  aluminum on  the inner side  only.  In the  center of
each straw  there is  a 20~$\mu$m diameter  sensing wire made  of gold
plated tungsten (W(Re)), stretched with  a tension of 40~g.  The wires
are aligned with a precision of $\pm 20\;\mu$m.

This  design  was chosen  in  order to  cope  with  the expected  high
particle  flux allowing  a maximum  deposited energy  of approximately
70~MeV/mm/s  for the  most exposed  straws at  the inner  part  of the
chamber.

The layers are  located between radii of 41 and  203~mm. The straws in
nine layers are parallel to the beam axis (z-axis) and the other eight
layers have  small skew  angles (6$^\circ$-9$^\circ$) with  respect to
the z-axis. These {\em stereo} layers form a hyperboloidal shape.

Due to  the forward cone  (Fig.~\ref{fig:mdcpipe}), the straws  in the
five inner  layers are positioned unsymmetrically with  respect to the
pellet pipe. The straws in each layer are inter - spaced by small gaps
in order to prevent the mechanical deformation by neighboring tubes.

The MDC is fitted inside a cylindrical cover made of 1~mm Al-Be and is
placed inside the solenoid (Fig.~\ref{fig:mdccyl}).

The  straws in  each (half)  layer are  mounted  between $\approx$5~mm
thick  Al-Be  end-plates.  The  layers  are assembled  around  the  Be
beam pipe  (Fig.~\ref{fig:mdcpipe})  and  the  attached pipe  for  the
pellets.  The beam pipe  has a diameter of 60~mm  and a wall thickness
of  1.2~mm.  

\begin{figure}[!htb]
\centerline{\includegraphics[angle=-90,width=0.6\textwidth,clip]
{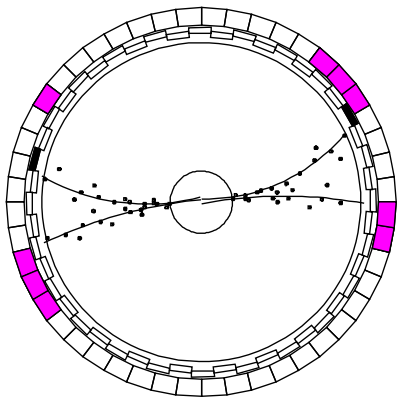}}
 \caption{Tracks in the MDC as seen in the event display of a 
          $\eta\to~e^+e^-e^+e^-$ decay candidate. 
 \label{fig:sed4e}
}
\end{figure}

Fig.~\ref{fig:sed4e}   shows   an  event   display   of  an   $\eta\to
e^+e^-e^+e^-$  decay  candidate.    The  lines  represent  the  tracks
reconstructed  in the MDC  from the  pattern recognition  program. The
apparent spread  of points is  due to hits  in the stereo  layers from
forward/backward  going tracks.   The  shaded areas  in the  outermost
rings gives the projection of the hit PSB central elements and the hit
SEC crystals (the size of the  crystals and the radial position of the
front faces are not to scale).

\subsubsection{The Plastic Scintillator Barrel - (PSB)}
The PSB is located inside the  SCS and surrounds the MDC.  It provides
fast signals for the first  level trigger logic and, together with the
mini drift chamber and the CsI calorimeter, it is employed for charged
particle identification  by the $\Delta E-p$ and  $\Delta E-E$ methods
and serves as a veto for $\gamma$ identification.

\begin{figure}[!hbt]
\centering
\framebox{
\includegraphics[width=0.275\textwidth,clip]{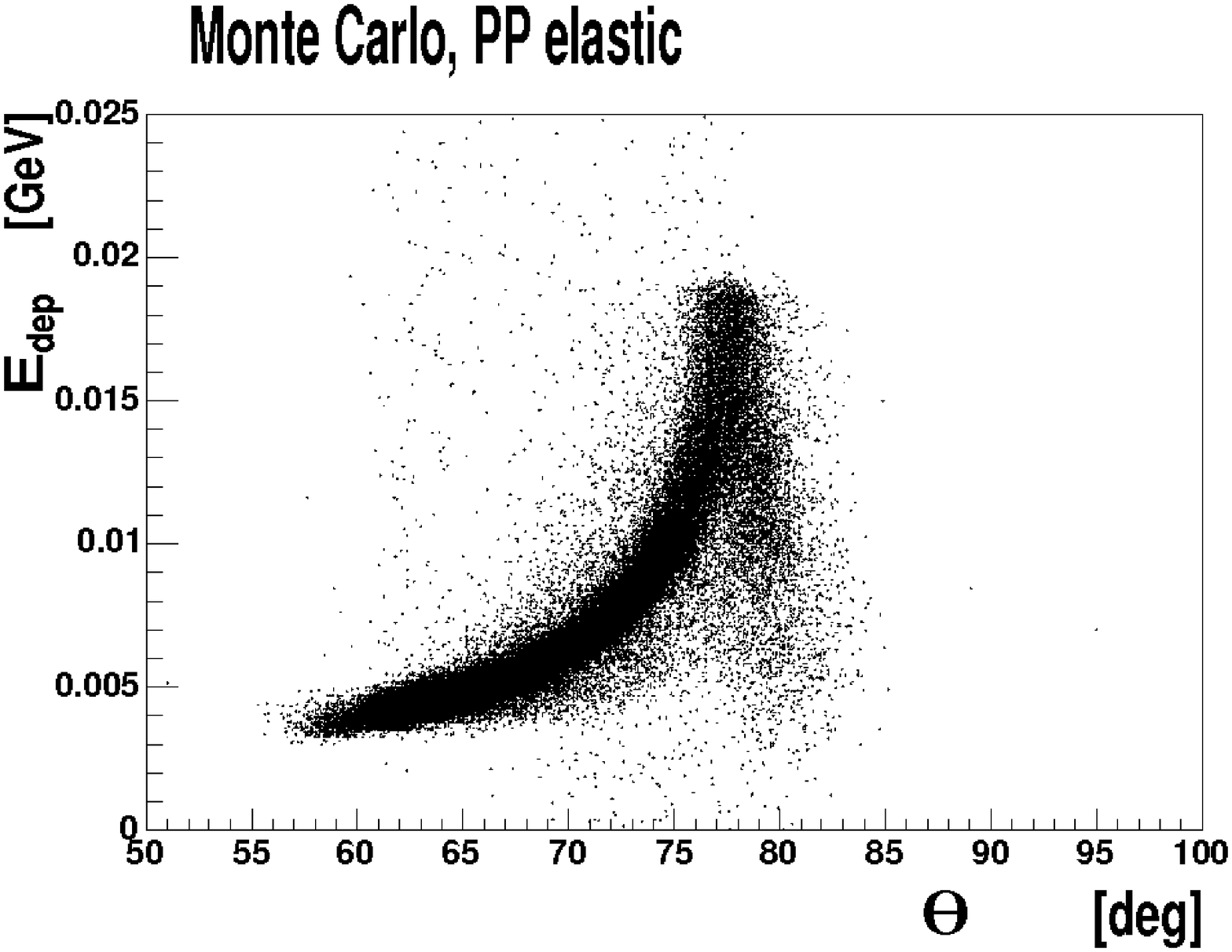}
}
\framebox{
\includegraphics[width=0.55\textwidth,clip]{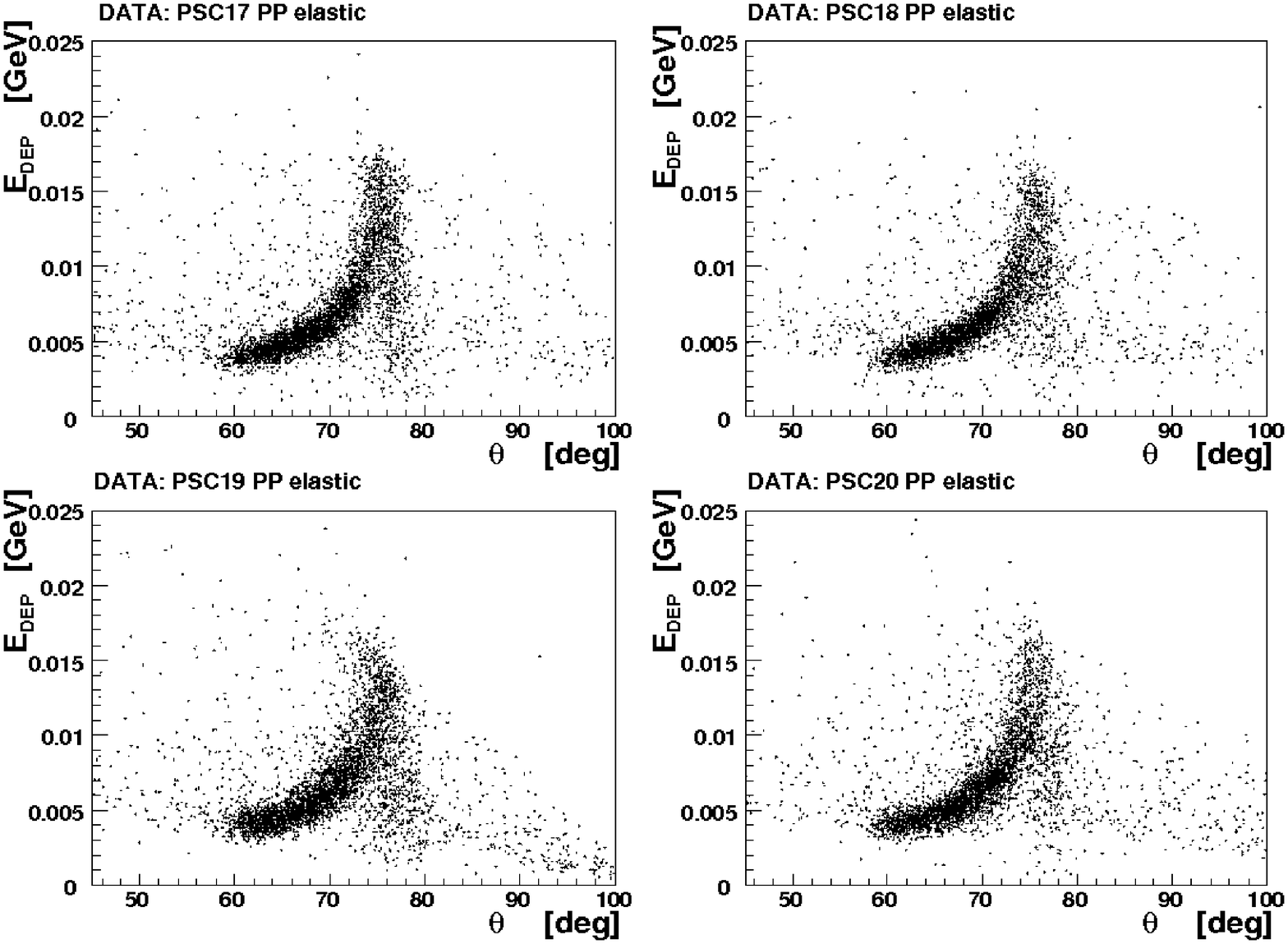}
}
\caption{\label{fig:pscal} (Left) Angular  dependence of the deposited
energy in  the PSB for  simulated elastically scattered  protons.  The
energy deposition increases with increasing polar angle (corresponding
to a  decrease of the kinetic  energy of the  proton), until particles
begin  to  stop  in  the  plastic  scintillator  material  (at  around
$\theta=77^\circ$).   (Right)  Experimental  spectra  corrected  for
light attenuation for four of the PSB central elements.}
\end{figure}

The  performance  of the  PSB  has  been  studied using  proton-proton
elastic scattering events.  Fig.~\ref{fig:pscal} (left plot) shows the
result of  a Monte Carlo simulation  of the angular  dependence on the
energy deposited in the  PSB. Fig.~\ref{fig:pscal} (right plots) shows
typical experimental spectra after  a correction for nonuniform signal
response has been applied.

\begin{figure}[!hbt]
\centering
\includegraphics[width=0.4\textwidth,angle=-90,clip]{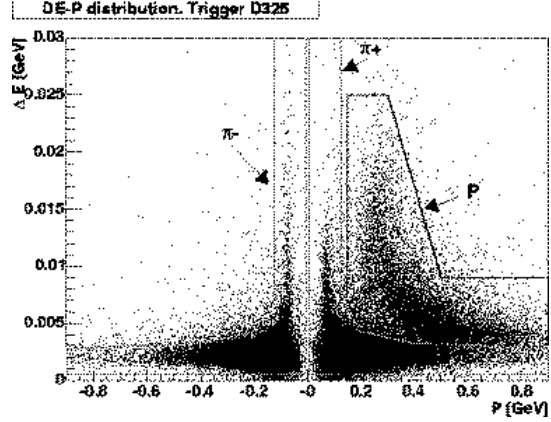}
\caption{\label{fig:CDdep} Example  of particle identification  in the
central detector for  a raw data sample collected  at 1360~MeV. Energy
deposited in  the PSB is given  as a function of  signed momentum from
the MDC. The regions for protons and pions are marked.  }
\end{figure}

In the initial experiments  the momentum and energy resolution allowed
reasonable  discrimination   between  pions  and   protons,  which  is
illustrated   in  Fig.~\ref{fig:CDdep}.    For  high   energy  charged
particles also  SEC information is available  and can be  used for the
identification.

\begin{figure}[!hbt]
\begin{minipage}[t]{0.35\linewidth}
\vspace{0.5cm}
\includegraphics[width=\textwidth,clip]{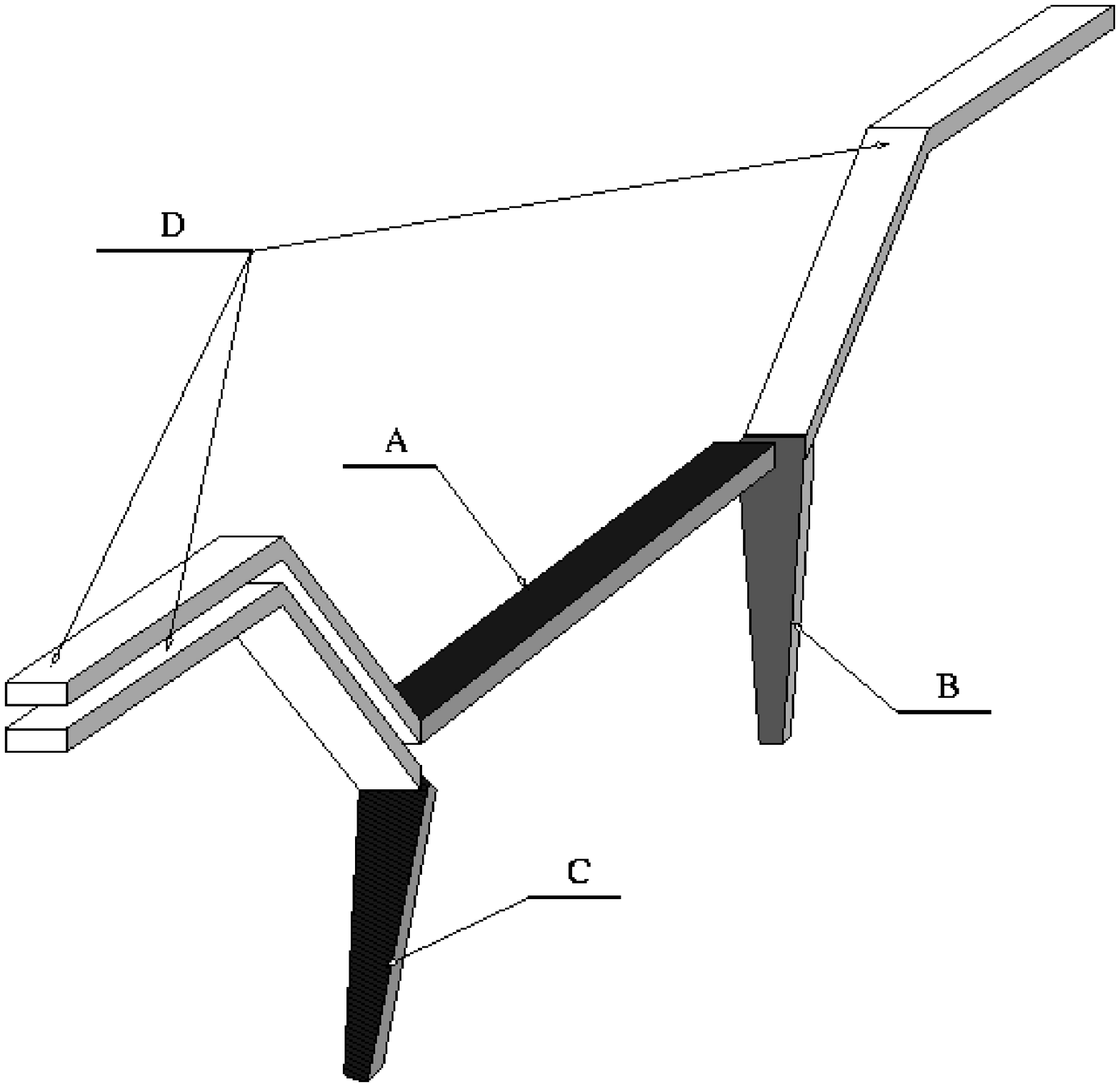}
\end{minipage}
\begin{minipage}[t]{0.65\linewidth}
\includegraphics[width=0.45\textwidth,angle=-90,clip]{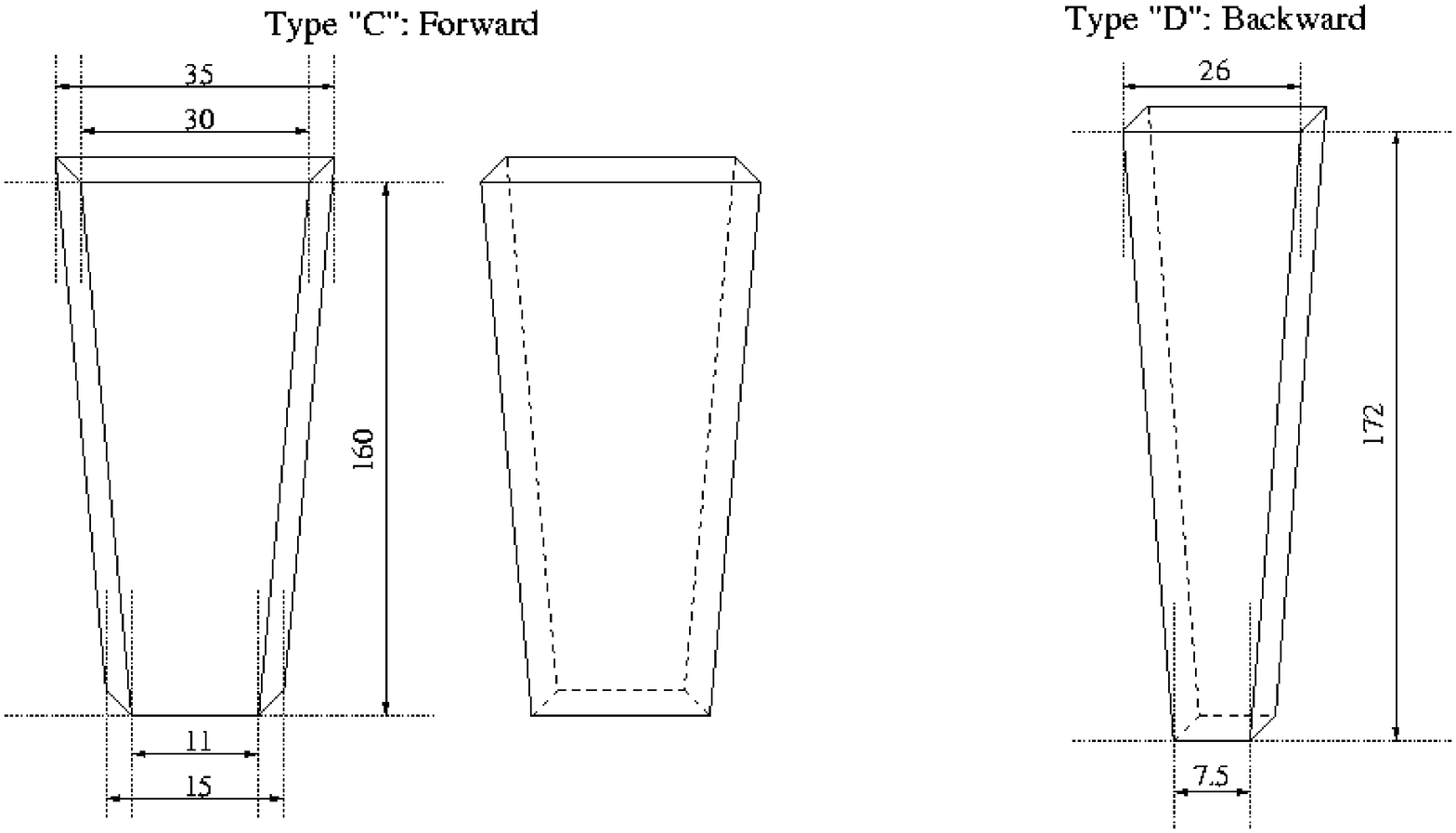}
\end{minipage}
\caption{\label{fig:psbsection} (Left) Layout of one section of the PSB
detector. {\bf A} denotes the rectangular counters of the barrel wall and
{\bf B } and {\bf C} are trapezoidal elements in the forward and in the
backward caps respectively. {\bf D} are bent light guides. (Middle)
Two shapes of the trapezoidal forward elements with dimensions marked
in mm. (Right) Shape and dimensions in mm of the trapezoidal backward
element.
}
\end{figure}

The PSB consists  of a cylindrical part and two  end caps and contains
in  total 146  elements shaped  as strips  of 8~mm  thickness.  In the
cylindrical part there are 48(+2)  elements of 550~mm length and 38~mm
width, forming 2 layers with a small (on average 6~mm) overlap between
neighboring   elements   to   avoid   that  particles   pass   without
registration.  The  end caps with  an outer diameter  of approximately
42~cm  in  the backward  and  51~cm in  the  forward  part contain  48
cake-piece shaped elements  each. The front end cap  is flat while
the rear  cap forms a  conical surface. Both  end caps have  a central
hole  for the  beam  pipe (19~cm  diameter  in the  forward and  12~cm
diameter in  the backward part).   One sector of  the PSB is  shown in
Fig.~\ref{fig:psbsection}.

Each scintillator  is glued to an  acrylic light guide  coupled to the
photomultiplier tube (PMT).   The PMTs are placed outside  of the iron
yoke  to  shield them  from  the  magnetic  field. For  this  purpose,
approximately 50~cm long light guides are used.

\subsubsection{The Scintillator Electromagnetic Calorimeter - (SEC)}
The  CD calorimeter  SEC is  able  to measure  photons, electrons  and
positrons  with  energies up  to  800~MeV.  The  energy threshold  for
detection of photons is about 2~MeV. The SEC consists of 1012 sodium-doped
CsI  scintillating   crystals  placed  between   the  superconducting
solenoid and the  iron yoke. The scattering angles  covered by the SEC
are between 20$^{\circ}$ and 169$^{\circ}$.

\begin{figure}
\begin{center}
\includegraphics[width=0.7\textwidth,height=0.4\textheight,clip]
{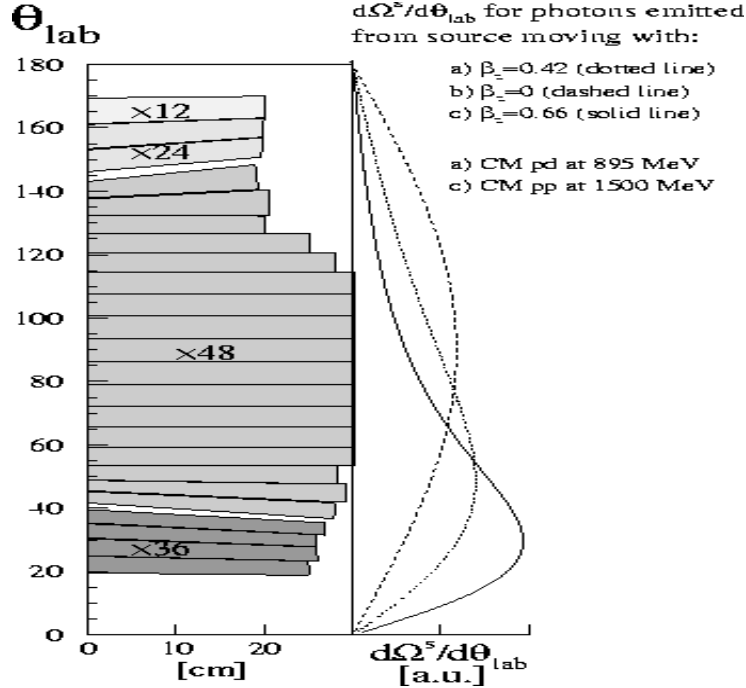}
\end{center}
\caption[The angular coverage  of the SEC] {\label{secross}The angular
coverage    of    the    SEC.     The   CM    system    solid    angle
($d\Omega^s/d\theta_{LAB}$)    vs.    the    LAB    scattering   angle
($\theta_{LAB}$) is  shown for $pp$ and $pd$  interactions at 1500~MeV
and 895~MeV.}
\end{figure}

\begin{figure}[htbp]
\centerline{\includegraphics[width=0.8\textwidth,clip]{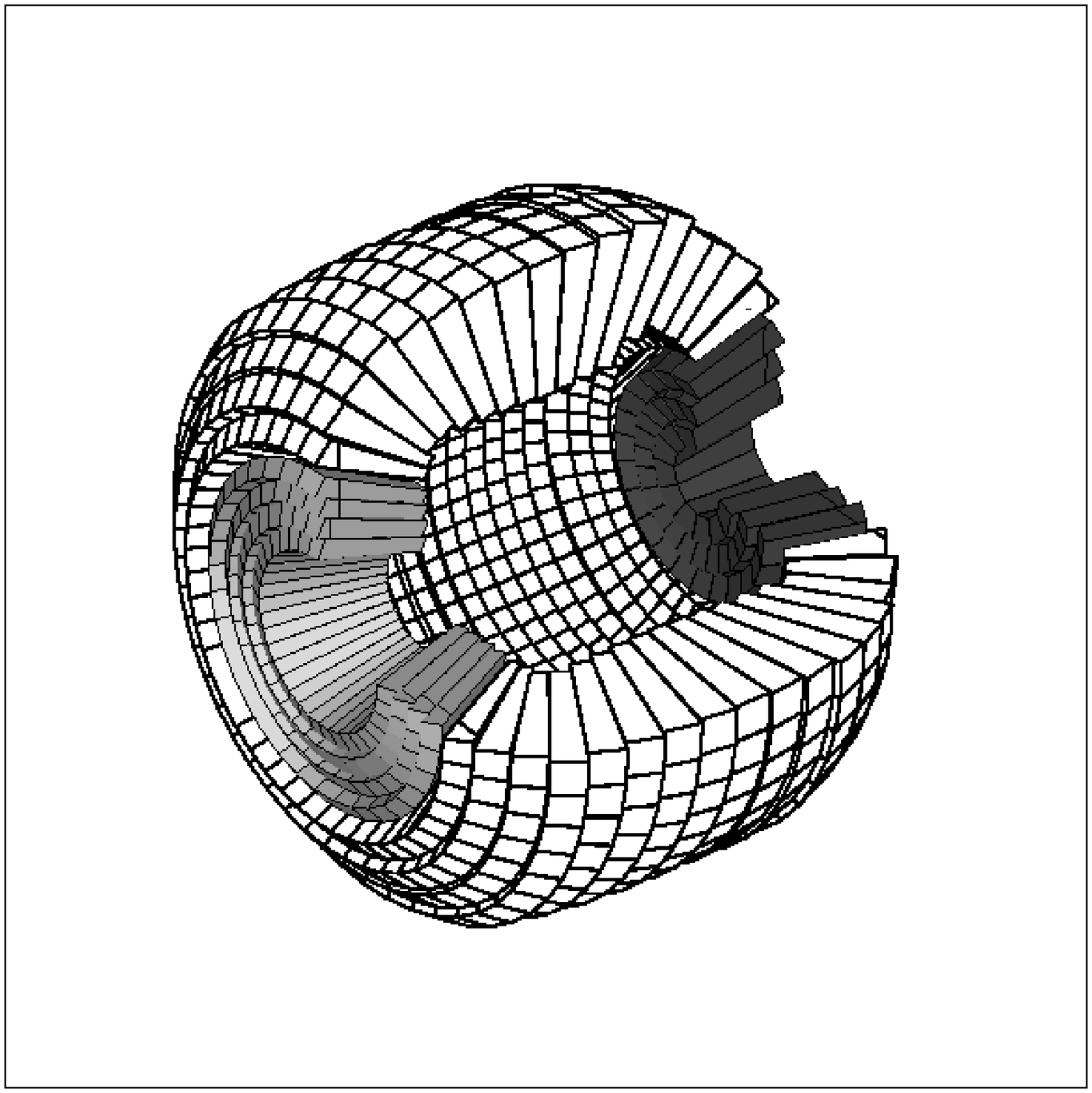}}
\caption[A schematic view of the SEC]
{\label{se}Schematic view of the SEC. It consists of the forward
part (shadowed
area to the left), the central part (not shadowed area in the middle)
and the backward part (shadowed area to the right). The beam is
coming from the right side.}
\vspace*{5mm}
\end{figure}

The crystals  are shaped  as truncated pyramids  and are placed  in 24
layers along the beam  (Fig.s~\ref{fig:wasa} and \ref{se}). The length
of  the crystals  varies from  30~cm (16.2  radiation lengths)  in the
central part to  25~cm in the forward and 20~cm  in the backward part.
Fig.~\ref{secross}  shows  the  angular  coverage  together  with  the
thickness of the SEC.  As  a measure of the anticipated photon fluxes,
the center of  mass (CM) system solid angle  vs.\,the laboratory (LAB)
scattering angle is shown for some experimental conditions at WASA.

The  forward part  consists of  4 layers  with 36  elements  each.  It
covers scattering angles from nearly 20$^{\circ}$ to 36$^{\circ}$. The
central part consists of 17 layers each having 48 elements, and covers
scattering  angles from 36$^{\circ}$  to 150$^{\circ}$.   The backward
part consists  of three layers.  Two  layers have 24  elements and the
layer  closest to the  CELSIUS beam  pipe has  only 12  elements.  The
small  spaces between the  forward-central and  central-backward parts
are  occupied by  PSB  light  guides and  mechanical  support for  the
solenoid (back  end only).  To  prevent photons and fast  particles to
escape  undetected,  these  spaces  (and  the  neighboring  layers  of
crystals)  are not  pointing  exactly towards  the interaction  region
(Fig.~\ref{secross}).  The calorimeter  covers nearly 360$^{\circ}$ in
azimuthal angle. Holes for the  pellet pipe (2+2 crystals) and for the
solenoid chimney (4 crystals) are not shown in the figure. Some design
parameters of the calorimeter are given in table~\ref{sec2}.

\begin{table}[bht]
\begin{center}
\begin{tabular}[b]{lr} \hline
Amount of sensitive material [g/cm$^2$] & 135 \\
\hspace*{3mm} [radiation lengths] & ${\rm \approx 16}$ \\
\hspace*{3mm} [nuclear interaction length] & ${\rm \approx 0.8}$ \\
Geometric acceptance: & 96\% \\
\hspace*{3mm} polar angle [degrees] &  ${\rm \approx 20 - 169}$\\
\hspace*{3mm} azimuth angle [degrees] & ${\rm \approx 0 - 360}$ \\ 
Max kinetic energy for stopping &  \\
\hspace*{3mm} ${\rm \pi^{\pm}}$/proton/deuteron & 190/400/500 \\
Scattering angle resolution [degrees] & ${\rm \approx 5}$ (FWHM) \\ 
Time resolution [ns] &  \\ 
\hspace*{3mm}charged particles & 5 (FWHM) \\ 
\hspace*{3mm}photons & ${\rm \approx}$40 (FWHM)\\ 
Energy resolution &  \\ 
\hspace*{3mm}charged particles  & ${\rm \approx 3\%}$ (FWHM) \\
\hspace*{3mm}photons  & ${\rm \approx 8\%}$ (FWHM) \\ \hline
\end{tabular}
\end{center}
\caption[SEC design parameters]
{\label{sec2}SEC design parameters.}
\end{table}

The SEC  is composed of sodium-doped CsI  scintillating crystals. This
type of scintillator material provides  a large light yield, has short
radiation length  and good mechanical properties.   CsI(Na) was chosen
instead  of  the more  commonly  used  CsI(Tl)  scintillators for  the
following reasons \cite{Ruber:1990aa,Schuberth:1995zz}:

\begin{itemize}
{\item  Its emission  peak at  420~nm matches  well the  bi-alkali S11
photocatode of  the selected PM  tubes, giving good  photon statistics
and sufficiently fast response.
\item Its shorter scintillation  decay time is preferable in high-rate
applications.
\item CsI(Na) gives much less afterglow than CsI(Tl).
\item  CsI(Na) seems  more  resistant against  radiation damage.  When
irradiated by a  proton beam corresponding to 10  years of operation a
test crystal  did not  show any visible  change in its  structure. The
CsI(Tl) test crystal, on the contrary, lost its transparency.}
\end{itemize}

\begin{figure}[htbp]
\begin{center}
\includegraphics[width=0.5\textwidth,clip]{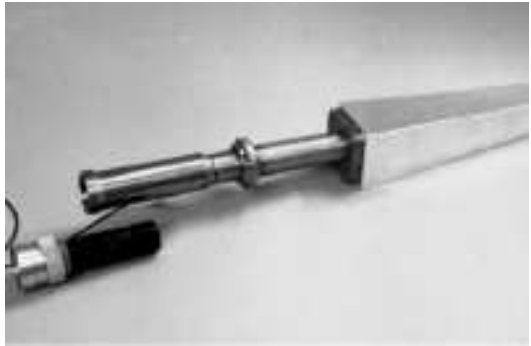}
\end{center}
\caption[A fully equipped CsI module]
{\label{csimodule}A fully equipped module with CsI crystal,
light guide, PM tube and housing.}
\end{figure}

The crystals are  connected by plastic light guides,  120~mm to 180~mm
long,  with the  photomultipliers placed  on the  outside of  the iron
yoke.   In Fig.~\ref{csimodule}, a  fully equipped  single calorimeter
module consisting  of a CsI  crystal, a light  guide, a PM tube  and a
high voltage unit, enclosed inside a special housing, is shown.

\begin{figure}[ht]
\includegraphics[width=0.49\textwidth,clip]{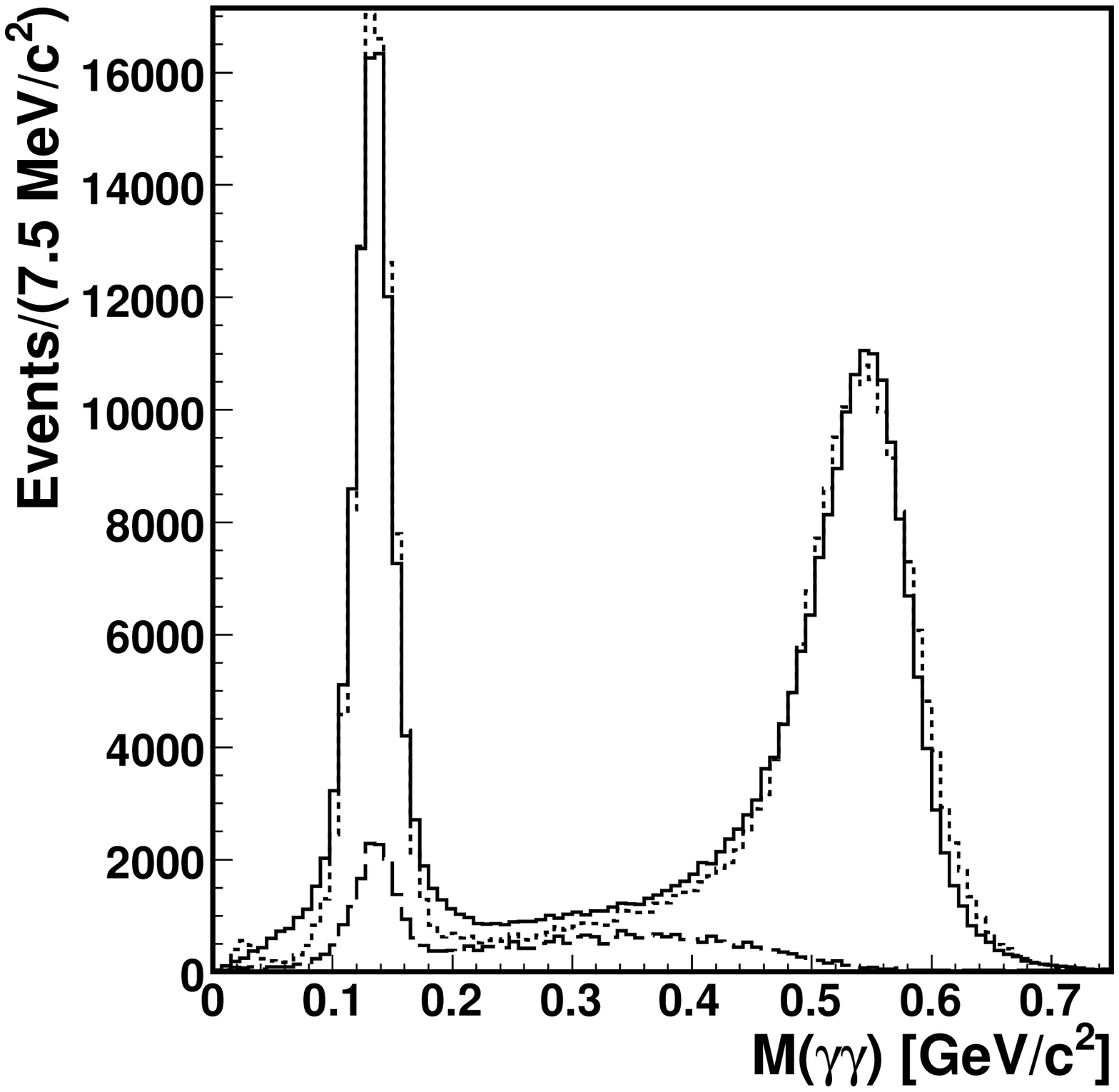}
\includegraphics[width=0.49\textwidth,clip]{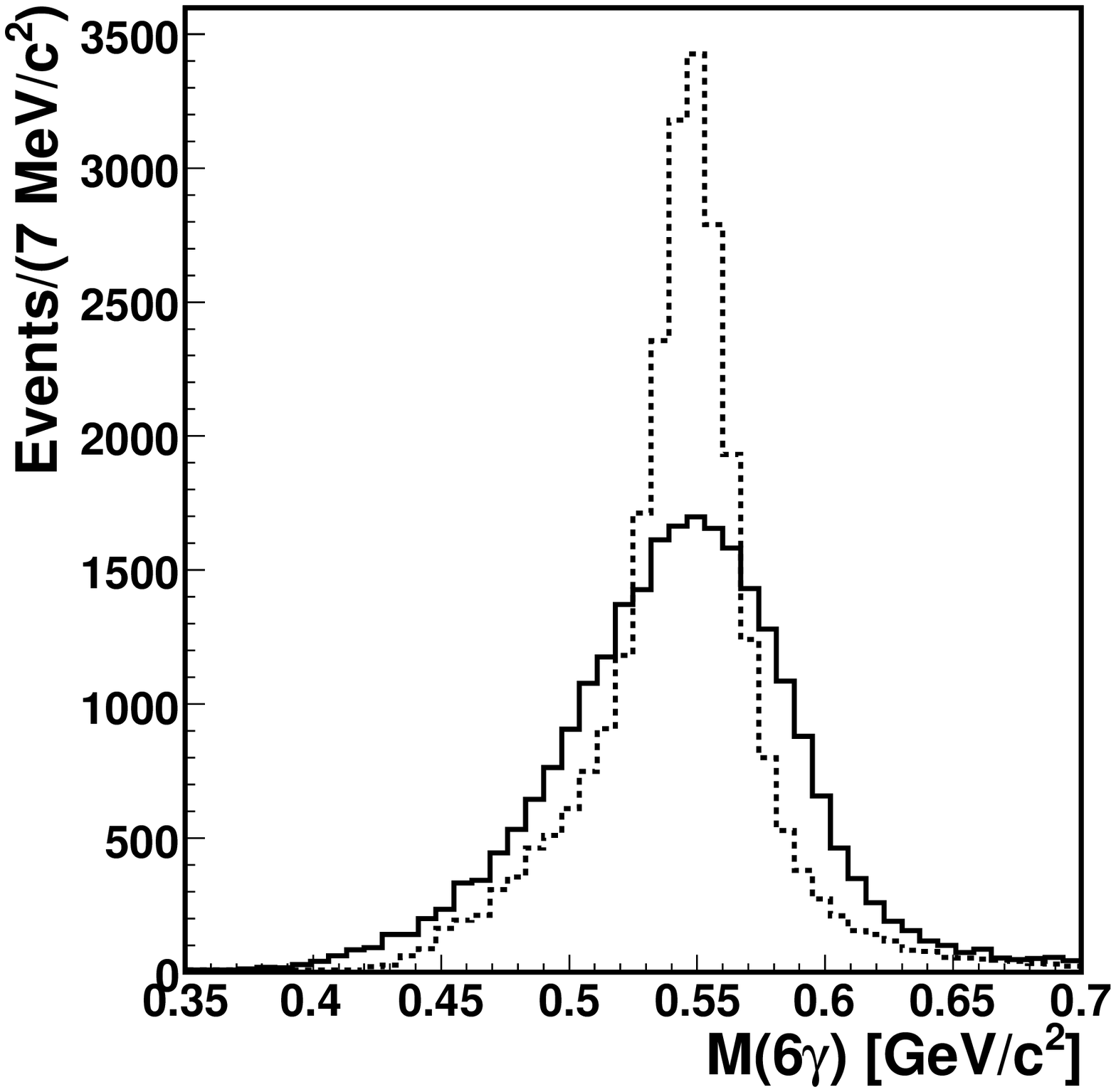}
\caption[Calorimeter  performance]  {\label{csiperf} (Left)  Solid line --
invariant mass  of two $\gamma$. Dotted line -- simulated data including 
the $\eta\to~2\gamma$ decay and direct $\pi^0$ and 2$\pi^0$ production.
Dashed line -- the contribution from 2$\pi^0$ production separately.
(Right)  Solid line - invariant  mass  of six
$\gamma$ (mainly from  $\eta\to~3\pi^0$ decay), dashed line --  after 
a kinematic fit constrained by
that  there  are  three  pairs  of photons  originating  from  $\pi^0$
decays.}
\end{figure}

The  performance of  SEC  for photon  measurements  is illustrated  in
Fig.~\ref{csiperf} by  data from proton-proton  collisions at 1450~MeV
beam proton kinetic  energy . Events with two  protons measured in the
FD with a missing mass in  the eta region are selected. The plots show
the invariant mass of the system of $\gamma$ for events with 2$\gamma$
and 6$\gamma$  detected. The energy  resolution in the  simulated data
(for  the  2$\gamma$ case)  corresponds  to  ${\rm \sigma_E/E  \approx
0.05/\surd  E(GeV)}$.  The  SEC is  described  in more  detail in  the
Ph.D. thesis of I.~Koch \cite{Koch:2005aa}.

\subsection{The Light Pulser System - (LPS)}

The  LPS delivers  reference  light  pulses via  light  fibers to  all
scintillation  counters in  order  to monitor  their  gain during  the
experiment. Since both slow and fast scintillators are used, two types
of light sources  were designed. A xenon flash  tube from Hamamatsu is
used for the CsI elements of the calorimeter and three LED-based light
sources,     of    the     type    used     in     the    PROMICE-WASA
setup~\cite{Zabierowski:1994mu},   supply  reference  pulses   to  all
plastic scintillators.  From those  four sources the light signals are
distributed to individual elements via a network of light fibers.  All
light sources  are monitored for stability, using  PIN photodiodes and
stable  amplifier circuits. The  relative resolution  of of  the light
pulse  signal amplitude  from the  plastic scintillators  is typically
1.5\%.    More   details   about    the   LPS   will   be   found   in
ref.~\cite{Zabierowski:2008aa}.

\subsection{Data Acquisition System - (DAQ)}
\label{subsubdaq}

\begin{figure}[hbt]
 \begin{center}
  \resizebox{0.95\textwidth}{!}{\includegraphics{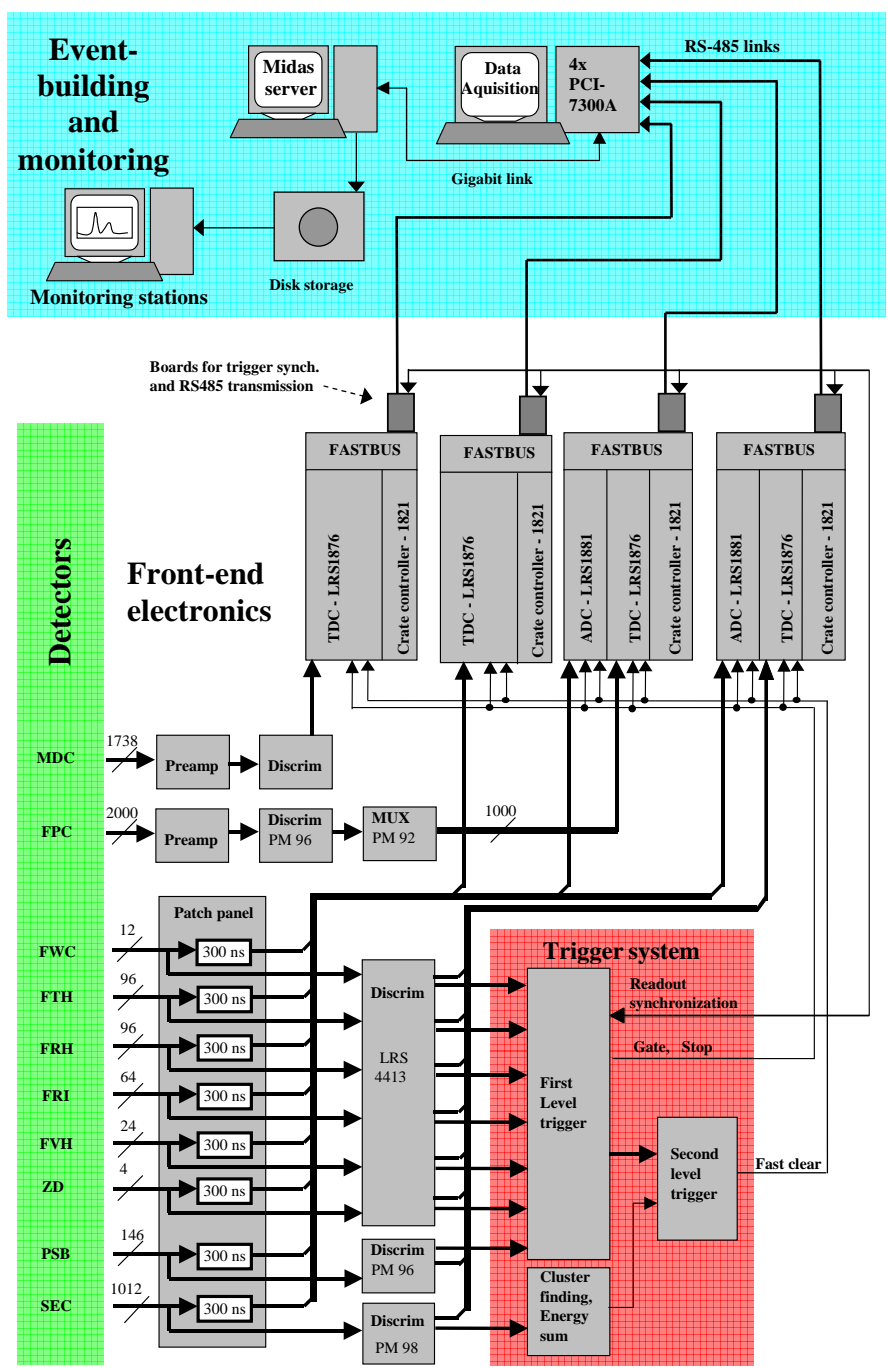}}
 \caption{Structure of the WASA DAQ and trigger 
          system~\cite{Fransson:2002ak,Marciniewski:2001zz}.}
 \label{fig:zel1} 
 \end{center}
\end{figure}

The DAQ  (Fig.~\ref{fig:zel1}) handles the signals from  the more than
5000   different  detector   elements  of   WASA.   All   signals  are
discriminated to  provide also time information  and logic information
for use in the trigger. The discriminators for the CD and the FPC were
developed  within the  collaboration while  the FD  scintillators used
commercially  available leading edge  discriminators. The  FPC signals
were "time multiplexed" after the discriminators by delaying every 2nd
channel and  combining the channels  2 and 2  to reduce the  number of
digitizing channels.

The  trigger system  is based  on specially  developed  hardware.  The
first  level  trigger  uses  discriminated signals  from  the  plastic
scintillators and generates gates and stops for the QDC's and TDC's. A
multiplicity is calculated for each detector plane, with clustering of
neighbors.   Also  several  consecutive  planes, FWC-FTH-FRH,  can  be
required  to  have matching  hits  in  $\phi$  before calculating  the
multiplicity. The conditions from the different planes are combined in
a 48  input coincidence  matrix. In the  second level,  implemented by
using  fast  clears of  the  TDC's and  QDC's,  the  signals from  the
calorimeter are used.  The analog  PMT signals from the SEC are summed
in groups of 4x4 or 3x4  and an energy threshold is applied. The logic
signals from the groups can  be combined with information from the PSB
to give  information on  the number of  hits from charged  and neutral
particles.   The  analog signal  can  be  further  summed to  allow  a
threshold on  the total  energy in the  SEC.  An additional  method to
search for clusters use  the discriminated signals from the individual
SEC  elements. Special  hardware search  for and  count the  number of
clusters  of  neighboring  elements   in  the  2d  SEC  matrix.   Many
parameters  of the  trigger, like  coincidence conditions  and delays,
could be  remotely controlled  via an I2C-bus  built into  the trigger
crates.

The  readout system  is based  on  digitizing modules  in the  FASTBUS
standard. The  charge from the ~1500  PMT signals are  digitized by 23
LRS1881\footnote{http://www.lecroy.com/lrs/}   QDC's   and  the   time
information from  the MDC, the  multiplexed FPC and  the discriminated
PMT  signals,  by  47  LRS1876  TDC's. In  addition  there  are  three
STR200\footnote{http://www.struck.de/}   32   channels   scalers   for
monitoring of  trigger rates.  (In total over  6000 readout channels).
With zero suppression in the QDC's  and TDC's this gives an event size
of a few kbytes. The  modules are distributed over four fastbus crates
which are  read in parallel. In  case of a trigger  each LRS1821 crate
controller send the data over  a proprietary RS485 link to a PCI-7300A
interface in the  readout PC.  An event number  for synchronization is
added         by        the         link         hardware.         The
PCI-7300A\footnote{http://www.adlinktech.com/} is a 32 bit digital I/O
interface with  a maximum data rate  of 80 Mbytes/s  A typical readout
time between 250 and 300~$\mu$s was achieved. The readout PC sends the
data from  the 4 PCI-7300A over  gigabit ethernet to  another PC where
the  data is  written  to a  disk  array and  monitoring stations  can
connect  to look  at the  data. The  subevents from  the 4  crates are
combined offline by the monitoring and analysis programs.

For the ZD internal detector setup, a slower separate data acquisition
system  was  used  in  parallel  to  read  the  Si  and  Ge  detectors
\cite{Hovander:1993aa}.   A threshold  on energy  in the  Ge detectors
provided a typical  trigger rate of a few Hz  which also triggered the
main  DAQ.  The main  DAQ  also recorded  the  event  number from  the
separate system so that the events could be correlated offline as well
as some Ge pulseheight information for cross check.

\subsection{Slow control}

A large number of detector  parameters in WASA required monitoring and
setup  independent of  the DAQ.   For  example the  pellet target  and
solenoid whose operation in addition required a close integration with
the operation of CELSIUS.  To facilitate this, their control, together
with  the high  voltage  for the  FD  PM tubes  and  the trigger,  was
integrated in the same system used to control the TSL accelerators and
beam-lines.   This  runs   on  VME   based  VxWorks   processors  each
communicating  with  a subset  of  the  hardware.  All processors  are
connected  together   in  a  dedicated  ethernet   network  with  UNIX
workstations  where graphics  interfaces  based on  Tcl/Tk and  SL-GMS
Synoptics\footnote{http://www.sl.com/}   allows   control   over   the
parameters.  The   parameters  are  stored   in  a  database   on  the
workstations with local  copies on each VME processor  of the relevant
subset.  All parameters  are logged  at regular  intervals as  well as
changes to the settings.

Also accelerator-cycle dependent quantities like trigger rates, pellet
rate and beam current were monitored  separately from the DAQ by 3 VME
scalers.           The         MIDAS          data         acquisition
system\footnote{http://midas.psi.ch/} was  used to log  the rates. The
web interface of MIDAS could be used to monitor the rates in real time
from any computer.

\section{Summary}
The 4$\pi$ detector facility  WASA, designed for studies of production
and decays  of light  mesons in proton  and deuteron reactions  at the
CELSIUS storage ring, has been presented. The facility is based on the
specially  developed pellet  internal  target system  which allowed  a
detector  coverage of  close  to 4$\pi$  steradians. Large  scattering
angles  are covered by  an electromagnetic  CsI calorimeter  and straw
chambers for  tracking of charged particles in  an solenoidal magnetic
field. At  forward angles  there are straw  chambers and  a multilayer
stack  of plastic scintillators.   He nuclei  escaping in  the CELSIUS
beam pipe can be detected in a zero-degree spectrometer.

\section{Acknowledgments}

We gratefully acknowledge the financial  support given by the Knut and
Alice  Wallenberg Foundation,  the  Swedish Council  for Planning  and
Coordination of  Research, the  Swedish Research Council,  the G\"oran
Gustafsson  Foundation,  the Swedish  Royal  Academy  of Science,  the
Swedish   Institute,   the   Swedish  Foundation   for   International
Cooperation in  Research and Higher Education, the  Polish Ministry of
Science and Higher Education PBS 7P-P6-2/07, the German BMBF (06HH152,
06HH952, 06TU261, 06TU201, 06TU987),  the Russian Foundation for Basic
Research (RFBR 02-02-16957), the Russian Academy of Science, the Joint
Institute for Nuclear Research in  Dubna, and the Japanese Ministry of
Education.   We also want  to thank  the technical  and administration
staff at The Svedberg Laboratory and at the participating institutes.

\bibliographystyle{elsart-num}
\bibliography{wasanim}

\end{document}